**Comparison of the X-TRACK Altimetry Estimated Currents with Moored ADCP and HF radar Observations on the West Florida Shelf**

Yonggang Liu[a*], Robert H. Weisberg[a],
Stefano Vignudelli[b], Laurent Roblou[c], and Clifford R. Merz[a]

[a]*College of Marine Science, University of South Florida, 140 7th Avenue South, St. Petersburg, Florida 33701, USA, yliu18@gmail.com; Weisberg@marine.usf.edu; cmerz@marine.usf.edu*

[b]*Consiglio Nazionale delle Ricerche, Area Ricerca CNR, Via Moruzzi 1, 56127, Pisa, Italy, vignudelli@pi.ibf.cnr.it*

[c]*Université de Toulouse; CNRS/CNES/IRD/UPS; Laboratoire d'Etudes en Géophysique et Océanographie Spatiales (LEGOS); 14 av Edouard Belin, F-31400 Toulouse, France, laurent.roblou@legos.obs-mip.fr*

(2011-08-08)

* Corresponding author email: yliu18@gmail.com, and phone # 1-727-553-3508

**Abstract**

The performance of coastal altimetry over a wide continental shelf is assessed using multiple-year ocean current observations by moored Acoustic Doppler Current Profilers (ADCP) and High-Frequency (HF) radar on the West Florida Shelf. Across-track, surface geostrophic velocity anomalies, derived from the X-TRACK along-track sea level anomalies are compared with the near surface current vector components from moored ADCP observations at mid shelf. The altimeter-derived velocity anomalies are also directly compared with the HF radar surface current vector radial components that are aligned perpendicular to the satellite track. Preliminary results indicate the potential usefulness of the along-track altimetry data in contributing to descriptions of the surface circulation on the West Florida Shelf and the challenges of such applications. On subtidal time scales, the root-mean-square difference (rmsd) between the estimated and the observed near surface velocity component anomalies is 8 – 11 cm/s, which is about the same magnitude as the standard deviations of the velocity components themselves. Adding a wind-driven Ekman velocity component generally helps to reduce the rmsd values.

# 1. Introduction

Satellite altimeters play an important role in ocean observations by providing repeated Sea Surface Height (SSH) information for the world's oceans (e.g., Fu and Chelton, 1984; Benveniste, 2011). Altimeter-derived Sea Level Anomaly (SLA) data are often used to infer surface geostrophic current anomalies, which are usually good approximations of surface current anomalies in deep ocean regions. Thus, altimetry-derived surface velocity products are used to study large-scale circulation in the world's oceans (e.g., Le Traon and Morrow, 2001; Lagerloef et al., 2003; Johnson et al., 2007) and its semi-enclosed seas (e.g., Cipollini et al., 2008; Liu et al., 2008; Alvera-Azcárate et al., 2009). However, such altimetry applications are mainly limited to such regions. In addition to common problems with altimetry data (e.g., Schlax and Chelton, 1994; Mitchum, 2000; Gaspar et al., 2001; Le Provost, 2001; Chambers et al., 2003; Mitchell et al., 2004; Mitchum et al., 2004; Ponte et al., 2007), there are intrinsic limitations to the altimetry data in shallow seas and near coastlines (e.g., Anzenhofer et al. 1999; Andersen 1999; Deng et al., 2002; Vignudelli et al., 2005; Roblou et al., 2011). Prior and existing satellite altimetry missions were not designed for applications in coastal oceans where the dominant dynamical processes have shorter temporal and spatial scales than those of the open ocean (Chelton, 2001).

However, in recent years, reprocessing of the along-track altimetry data have been pursued for the purpose of improving the data quality within the coastal oceans (e.g., Vignudelli et al., 2005; Roblou et al., 2007; Madsen et al., 2007; Brown, 2010). As a result the useful altimetry data domain now extends further toward the coast. For example, with specialized corrections and processing strategies, Vignudelli et al. (2005) processed a SLA data set in a 30-km narrow path between Corsica and Capraia Islands, within the Corsica Channel. In the same area, a multi-mission data set was extended to a mean distance of 32 km from the coast (Bouffard et al., 2008). Despite innovative efforts from Dussurget et al. (2011), it is still challenging to get two-dimensional gridded SSH maps for coastal applications. The repeated along-track altimeter data is an independent source of oceanographic data that can provide useful information on coastal ocean dynamics. For example, the surface geostrophic velocity estimated from the along-track altimetry data may be used to study coastal ocean circulation on continental shelves, especially when the satellite tracks are in across-shelf direction (e.g., Han 2004b, 2006; Durand et al., 2008, 2009; Le Hénaff et al., 2011), and time series of surface geostrophic velocity vectors may be obtained at the crossover points of two intersecting satellite tracks (e.g., Bouffard et al., 2008).

Advances of coastal altimetry techniques and applications are summarized in a recent book (Vignudelli et al., 2011). Included is a review of coastal altimetry applications in North American coastal oceans by Emery et al. (2011). Applications of altimeter-derived currents were mostly on seasonal and longer time scales (e.g., Strub et al., 1997; Foreman et al., 1998; Strub and James, 2000; Li and Clarke, 2005; Venegas et al., 2008) and focused on currents on continental slopes (e.g., Kelly and Gille, 1990; Kelly et al., 1998; Dong and Kelly, 2003; Han, 2004a, 2007) or even more offshore (e.g., Leben and Born, 1993; Sturges and Leben, 2000; Leben, 2005). Recent coastal altimetry applications show that altimeter-derived currents may be extended to the coast by combining with tide gauges (e.g., Saraceno et al., 2008). However, such efforts are mainly for narrow continental shelves where the deep ocean influence is close to the tide gauges, allowing the SSH on the shelf to be linearly interpolated between the open ocean and coastal tide gauges.

In contrast, applications of altimeter-derived velocity are rarely seen over wide and shallow continental shelves (e.g., Han et al., 1993, 2002, Volkov et al., 2007). However, satellite altimetry showed some promise on the wide West Florida Shelf (WFS) (e.g., Liu and Weisberg, 2007). Based on dominant inner-shelf momentum balance, Liu and Weisberg (2007) proposed a method for estimating absolute SSH near the coast by integrating sea level inferences from in situ coastal ocean observations (time series of velocity, hydrography, bottom pressure, coastal tide gauge and winds) along a WFS shore normal transect. Compared with the satellite SLA from the merged AVISO product (e.g., Ducet et al., 2000; Le Traon et al., 2003; Pascual et al., 2006), the estimated SSH time series at the 50 m isobath site showed encouraging results (Liu and Weisberg, 2007). Recently, Carlson and Clarke (2009) estimated the monthly mean along-shelf component of the geostrophic velocity on the outer WFS, but without testing against in situ current observations.

Evaluation of altimeter-derived surface velocity was seen in many deep ocean studies (e.g., Menkes et al., 1995; Stammer, 1997; Strub et al., 1997; Dong and Kelly, 2003). Strub et al. (1997) evaluated TOPEX altimeter-derived velocities with moored Acoustic Doppler Current Profilers (ADCP) (at 4800 m water sites) velocities at depths below the Ekman layer for the California Current System. Dong and Kelly (2003) compared the altimeter-derived geostrophic currents with moored current meter records at 1000 – 3000 m water sites in the Middle Atlantic Bight. Recently, such validations were found in coastal oceans. For example, Powell et al. (2006) compared TOPEX/Poseidon (T/P) and Jason-1 (J1) altimeter-derived velocity time series with moored ADCP currents on the shelf in the northwestern Gulf of Mexico during a six-month period. Saraceno et al. (2008) evaluated a coastal SSH product of combined AVISO and tide gauge data sets by comparing the SSH-derived coastal velocity with both moored ADCP and HF radar currents on the narrow Oregon-California shelf. Vignudelli et al. (2005) and Bouffard et al. (2008) using respectively a single mission and a multi-mission configuration also evaluated altimeter-derived velocity data with moored currents over a coastal zone of the northwestern Mediterranean where the coastline and bathymetry are complex.

As a follow-on study of Liu and Weisberg (2007), this paper assesses the usefulness of satellite altimetry on a wide continental shelf with relatively simple bathymetry and coastline shape. Altimeter-derived geostrophic velocity anomaly time series are compared with the multiple-year velocity component records from moored ADCPs and HF radars on the WFS. The purpose is to determine to what extent the shelf currents may be approximated by altimeter-derived currents on subtidal time scales in the vicinity of the 50 m isobath or approximately near mid-shelf. This is one step toward future applications of coastal altimetry; for instance, to provide surface geostrophic current maps or to examine Loop Current and Loop Current eddy interactions with the shelf.

Data and data processing are described in Section 2. The methodology for estimating surface geostrophic velocity components is provided in Section 3. Velocity component comparisons, based on moored ADCP and HF radar data, are given in Sections 4 and 5, respectively. Section 6 provides a summary and discussion.

**2. Data**

*2.1 X-TRACK SLA altimetry product*

The altimetry data used in this analysis is based on Geophysical Data Records (GDR) provided by AVISO (AVISO 1996, 2008) for both the T/P and J1 missions. A coastal altimetry SLA product is then generated from the X-TRACK post-processing tool (Roblou et al., 2007, 2011) developed at Centre de Topographie des Océans et de l'Hydrosphère (CTOH). In particular, the CTOH/X-TRACK SLA data are subjected to the following corrections:

$$SLA = SLA0 - DAC - TIDES \tag{1}$$

where SLA0 is the X-TRACK SLA without applying the geophysical corrections, DAC includes the atmospheric loading effects, i.e., MOG2D-G model (Carrère and Lyard, 2003) sea level, and TIDES includes solid earth (Cartwright and Taylor, 1971), loading and ocean tides GOT4.7 (Ray, 1999; 2008). More details about the X-TRACK post-processing can be seen in Roblou et al. (2007, 2011). Recent applications of X-TRACK altimetry data were seen in the Mediterranean Sea (Birol et al., 2010; Bouffard et al., 2011), along the Spanish coasts (Le Hénaff et al., 2011; Dussurget et al., 2011; Herbert et al., 2011), and the Solomon Sea (Melet et al., 2010).

The satellite ground track 091 of T/P and J1 missions crosses the northern part of the WFS (Fig. 1a), and the data are used to compare with the moored ADCP data in this paper. The other track (167) passes by a more complicated topography area (parallel to and then crossing the isobaths) on the southern WFS, and it is not used in this analysis. Two satellite ground tracks (102 & 167) of T/P interlaced orbit passing over the central WFS are used to compare with the HF radar data (Fig. 1b). The 1 Hz along-track SLA data contain high-frequency gravity waves (e.g., Ray and Mitchum, 1996; Zhao et al., 2010) that are not of our interest at present. A 30 km 'box-car" low-pass filter is applied to each cycle of the track. This hopefully removes the high-wavenumber variations (the variance is reduced by about 5%). The distance between two adjacent points along the satellite track is about 6 km.

*2.2 In situ observations on the West Florida Shelf*

Velocity measurements by moored ADCPs began on the WFS in 1993, first at a single location (PM1) on the 47 m isobath (Fig. 1a), and then at multiple locations across the shelf (e.g., Weisberg et al., 1996, 2000, 2005, 2009a,b; Meyers et al., 2001; Liu and Weisberg, 2005a,b, 2007). Unfortunately, no moorings are located on any of the satellite ground tracks. Also, many of the moorings are located landward of the 30 m isobath, where the altimeter data are problematic. Thus, data from the moorings deployed around the 50 m isobath (PM1, C12 & C13) are chosen for a preliminary assessment of the altimeter-derived currents. Note that these locations are at least 90 km away from the coastline. With downward-looking ADCPs, currents measured across most of the water column (excluding roughly 4 m from either the surface or the bottom) are archived hourly. Quality control and editing procedures are provided by Liu and Weisberg, 2005a,b and Mayer et al., 2007).

Complementing the moored ADCPs, CODAR SeaSonde, long-range HF radars have been deployed on the WFS since September 2003 (Liu et al., 2010). These HF radars operate at a nominal frequency of 5 MHz, with the intended purpose of observing surface currents out to a range of about 200 km offshore. However, due to the low energy (waves and currents) nature of the WFS, the HF radar data returns are gappy in both space and time (Liu et al., 2010). When acquired, the velocity components by HF radar compare well with those by moored ADCPs. Radial data coverage fluctuates with time on different time scales, and varies from station to station (Liu et al., 2010). Among the three radar sites, the Venice site (27º 04.655′ N, 82º 27.096′ W) has the best data return, and thus its data are used in this analysis. Its radar spatial coverage/footprint is shown in Fig 1b. With such 5-MHz radar, the radial velocity estimates are at an effective depth of 2 m, with nominal range and bearing resolutions of 5.8 km and 5º, respectively. Hourly radial current maps are archived and processed using the CODAR SeaSonde software suite after directional calibration using measured antenna patterns as in Kohut and Glenn (2003). A community toolbox "HFR_Progs" is used to remove the outliers and to interpolate the radial data onto uniform radial grids. Detailed information about the HF radar data processing and the availability of the data during 2003-2008 are provided in Liu et al. (2010).

*2.3 Wind data and Ekman component computation*

Hourly wind data were obtained from the National Data Buoy Center (NDBC) buoy 42036 (online at http://www.ndbc.noaa.gov/station_page.php?station=42036) at the 55-m isobath (Fig. 1). The anemometer height for the buoy is 4 m. The hourly winds were adjusted to a standard 10-m level prior to further analysis. The wind stress is calculated using a variable drag coefficient depending on the magnitude of the wind speed (Blanton et al., 1989). Low-pass filtering with a 36 hour cut-off period provides wind stress time series for estimating the surface Ekman velocity component based on the deep water Ekman equations (Ekman, 1905; Lenn and Chereskin, 2009) with a constant eddy viscosity ($K = 0.02$ m$^2$/s).

## 3. Estimation of geostrophic velocity from along-track altimeter data

Internally generated white noise in altimetry data can affect the accuracy of the slope estimate from along-track SLA. It can dominate the errors in estimating mesoscale currents in coastal oceans because a simple finite-difference of the along-track SLA acts as a high-pass filter (Powell and Leben, 2004). This effect may be mitigated by using longer difference operator lengths, e.g., differencing over larger point-to-point intervals, which is equivalent to a running average of the geostrophic velocity component estimates along the satellite ground tracks (e.g., Strub et al., 1997). Powell and Leben (2004) provide a compact, optimal difference operator to minimize the white noise of the SLA measurements when computing SSH slope (and hence geostrophic velocity) by weighted smoothing along-track SLA:

$$\Delta_{Nt}^{pq} = \sum_{\substack{n=-p \\ n \neq 0}}^{q} c_n \left( \frac{h_{i+n} - h_i}{n \Delta t} \right), \tag{2}$$

where $h_i$ is the SLA at the current ($i$) point along the track, and $\Delta t$ is the sampling interval in seconds, $p$ and $q$ are the number of points before and after the current point, $N = p + q$, and $C_n$ are the weighting coefficients that satisfy

$$\sum_{\substack{n=-p \\ n \neq 0}}^{q} c_n = 1. \tag{3}$$

[We note a minor typo in Eq. (3) of Powell and Leben (2004), where the subscript $i+1$ should be $i+n$ as appeared in Eq. (2) of this paper]. The weights may be determined by minimizing the noise in calculating the slope of the along-track SSH. This optimal filter is ideal for coastal ocean studies where the baroclinic Rossby radius of deformation is small. Use of this filter may be found in Leben and Powell (2003), Powell et al. (2006), Deng et al. (2008), and Godin et al. (2009).

The optimal filter of Powell and Leben (2004) is used here to smooth the along-track sea level slope $\Delta h$, which is further used to compute across-track geostrophic velocity anomaly:

$$v_{geo} = \frac{g}{f} \frac{\Delta h}{\Delta x} \tag{4}$$

where $g$ is acceleration of gravity, $f$ is Coriolis parameter, and $\Delta x$ is the along-track distance between two adjacent points. A 12-point filter ($p = 5$, $q = 6$) is used here, and the length of 11 consecutive points along the satellite ground track is about 60 km, which is about twice the baroclinic Rossby radius of deformation (estimated, for instance, by He and Weisberg, 2003). The selection of the filter size is within the suggested range for this area, according to Figure 3 of Powell and Leben (2004). It is expected that this filter size results in a slope noise standard deviation of 4~6 cm/s in the geostrophic velocity (Powell and Leben, 2004).

**4. Comparison with moored ADCP observations**

*4.1. Mooring PM1*

Velocity data at mooring PM1 is available from October 1993 to January 1995. The corresponding satellite altimeter data is from T/P mission, which spans November 1992 to August 2002. The altimeter-derived surface geostrophic velocity component anomalies are compared to ADCP near-surface velocity component anomalies in two directions: (1) across-track direction (i.e., perpendicular to satellite ground track 091, Points N and PM1), and (2) along-shelf direction (Points P and PM1). The distance between point N and mooring PM1 is about 44 km (Fig. 1a). To get an estimate of the along-shelf velocity component at Point P, the across-track geostrophic velocity component at Point P is divided by $\cos(\theta)$, where $\theta$ is the angle between across-track direction and the along-shelf direction. Essentially, it is an exemplification of across-track velocity component at that point. This is consistent with the fact that the subtidal principal axis currents are in the along-shelf direction (Liu and Weisberg, 2005b). The near surface velocity vectors (with the top bin at 4 m below surface) are 36-hour low-pass filtered and then rotated to get the across-track or along-shelf velocity component. A record-long mean value is further removed to get velocity anomaly $v_{adcp}$. The altimeter-derived surface geostrophic velocity component anomalies ($v_{geo}$) are averaged among three adjacent points to further reduce the effects of small-scale features.

To examine whether a wind-driven surface Ekman velocity contribution plays a role, the subsurface (4 m) Ekman velocity is rotated to the across-track or along-shelf directions, respectively, and the mean value is removed for comparison ($v_{Ekm}$). Time series of across-track velocity component anomalies and their differences are shown in Fig. 2. The standard deviations of the moored ADCP $v_{adcp}$, altimeter-derived $v_{geo}$ and geostrophic plus Ekman velocity components ($v_{geo}+ v_{Ekm}$), at the altimeter data time stamps, are about the same (7.0, 8.1 and 8.8 cm/s, respectively). The root-mean-square difference (rmsd) between the altimeter-derived and moored velocity anomalies is 8.6 cm/s, about the same magnitude as the standard deviations of the velocity anomalies (Table 1). When the Ekman velocity component is added to altimeter-derived velocity anomaly, we see a slight reduction in the rmsd value. The residual time series is shown in Figure 2b. In a perfect comparison, the residuals would all be zero. We can see reductions of the residuals at some times (e.g., the two points in October 1993), however, there are still large variations of the residuals. The standard deviation of the residuals is 8.0 cm/s (7.9 cm/s with wind effect considered), which is about the same magnitude as the standard deviations of the velocities themselves. Results from Vignudelli et al. (2005), Saraceno et al. (2008) and Le Hénaff et al. (2010) also highlight an rms-difference comparable to the temporal velocity anomaly variability itself.

In the along-shelf direction (Fig. 3), the standard deviations of the velocity components, their rmsd values, and the standard deviations of the velocity differences are all slightly larger than those in the across-track direction (Table 1). This is consistent with the fact that subtidal currents on the WFS have the largest variability in the along-shelf direction as shown by the principal axis direction (e.g., Liu and Weisberg, 2005a,b; Weisberg et al., 2009b). Similar to those in the across-track direction, the rmsd of the velocity components and the standard deviation of the velocity difference are slightly larger than the standard deviations of the velocity time series. Adding the Ekman component also reduces the rmsd value.

*4.2. Moorings C12 and C13*

On subtidal time scales, and by virtue of the Taylor-Proudman theorem, the principal component of the subtidal velocity variations tend to align with the isobaths (e.g., Liu and Weisberg, 2005a,b; Weisberg et al., 2009b), and the decorrelation scales tend to be larger in the along-shelf direction than in the across-shelf direction. This provides a basis for comparing the along-shelf currents at two points that are on the same isobath but not far away from each other. Note that the 50 m isobath is fairly "Straight" on the WFS. Our next comparison is the altimeter-derived currents at point P and moored ADCP currents at C12 and C13, all in the along-shelf direction.

Altimeter data from J1 mission is used for these comparisons. J1 satellite has the same ground tracks as T/P (Fig. 1), and track # 091 is used again. J1 altimeter started data collection in January 2002. The along-track SLA data used here are inclusive of 2002 to 2007. The near-surface bin (5 m below the surface) ADCP data is used for comparison. As shown from Figs 5 & 6, there are over five years' continuous moored ADCP data at moorings C12 and C13, respectively, during the six-year period.

The distance between point P and mooring C12 along the 50m isobath is about 113 km. A time-lagged correlation analysis is performed between the altimetry-derived and observed velocity anomalies. Both the correlation coefficient ($\gamma^2$) and the rmsd values between the two velocity anomalies are calculated for the time lag between -40 and 40 hours (Fig. 4). The maximum correlation is found at 21 hour time lag, i.e., the observed velocity at mooring C12 lags the altimetry-derived velocity at track #091 about the 50 m isobath. The minimum rmsd value is found at 18 hour time lag, which is close to the 21 hour time lag obtained from the maximum correlation coefficient. Adding wind effect only change the time lag by 1~3 hours. Thus, the 18 hour time lag seems to be consistent from both the maximum correlation coefficient and minimum rmsd criteria. This time lag indicates an along-shelf propagation velocity of 1.7 m/s (113 km in 18 hours), which is consistent with a long-term mean along-shelf velocity or 2~3 cm/s along the 50 m isobath on the WFS (Weisberg et al. 2009b) Unfortunately, a consistent time lag is not found using the velocity at mooring C13, probably due to the fact the correlation coefficient between the altimetry-derived and observed velocity at C13 is not significantly high ($\gamma^2$ <0.1). Although a consistent time lag is not found as for C12, the rmsd values are within 1.5 cm/s difference for different time lags, which is the same as that from C12.

Similar to those for the along-shelf current comparison at PM1, the standard deviations of the altimeter-derived and moored ADCP along-shelf velocity anomalies are about the same (8.4~9.9 cm/s), and the rmsd between the estimated and the observed velocities is 9.9 cm/s at zero time lag (9.0 cm/s at 18 hour time lag). Adding subsurface (5-m) Ekman velocity anomaly to the altimeter-derived geostrophic velocity slightly increases the standard deviation of the velocity component from 9.9 cm/s to 11.6 cm/s. However, the addition of the wind Ekman component decreases both the rmsd value between the estimated and the observed velocities and the standard deviation of the estimated and the observed velocity difference (Table 1). That is to say, the comparison is improved by adding the Ekman component. Comparison between the estimated velocity at point P and the observed velocity mooring C13 further down south (Fig. 6) gives about the same results as those with mooring C12.

**5. Comparison with HF radar current observations**

All the above comparisons with ADCPs need velocity rotation/projection, either projecting the observed velocity to the direction perpendicular to the satellite track or rotating both estimated and observed velocities to the along-shelf direction. HF radar mapped radial velocities (Fig. 7) offer a direct velocity comparison without velocity rotation if satellite ground tracks pass over the radar coverage area. There is a point Q along the track that the HF radar radial current direction is perpendicular to the satellite ground track (Fig. 1b). Since altimeter-derived geostrophic velocity is also in the across-track direction (line Q – Veni), the two velocity components can be directly compared. Also, the effective depth of HF radar current measurements is the top 1 – 2 m, which is more comparable to the altimeter-derived surface geostrophic velocity than the moored ADCP velocity top bins (4 – 5 m below surface).

Satellite ground tracks 102 & 167 of T/P interlaced orbit cross over the footprint of Venice HF radar on the central WFS (Fig. 1b). Velocity comparisons are made at two points, Q and R, respectively. An average of the altimeter-derived surface geostrophic velocities at three points centered at point Q (or R) is applied to obtain $v_{geo}$ time series. Similarly, an average of the HF radar radial velocities at the sectors at and around point Q (or R) is used as $v_{rad}$ time series. The spatial ranges of the radial data points used for averages are shown in Fig. 1b as the small boxes, which contain three sectors in bearing and radial directions, respectively. The spatial average over adjacent sectors helps to retain more data, which is necessary for the gappy data (Liu et al., 2010); however, it might also bring in uncertainties. After 36 hour low-pass filtering, not much HF radar data are left. So, we opt with a shorter cut-off (30 hour) and retain more data points in the $v_{rad}$ time series. Comparison is performed on a period with good HF radar data return during 2004 - 2005.

Point Q is located on the inner shelf (Fig. 1b & Fig. 7), closer to the coast than the points (P & R) around the 50 m isobath, and the standard deviation of the altimeter-derived surface geostrophic velocity is slightly larger than those at point P. The standard deviation is reduced to 9.3 cm/s when surface Ekman velocity component is taken into account (Fig. 8), but it is still much larger than the standard deviation of the HF radar radial velocity (4.9 cm/s). Similar to those of moored ADCP comparisons, Ekman velocity component helps to reduce the rmsd between $v_{geo}$ and $v_{rad}$ and the standard deviation of the velocity difference.

Point R is located on the mid shelf (Fig. 1b & Fig. 7), and the standard deviation of the HF radar radial currents is larger than that at inner shelf point Q (7.9 vs. 4.9 cm/s). This is because surface currents have larger variability (Fig. 7), and the ellipses of the principal axis currents are less eccentric, on the mid to outer shelf than on the inner shelf (e.g., Liu and Weisberg, 2005a; Weisberg et al., 2009b). The standard deviation of $v_{geo}$ is smaller than that at point Q (3.8 vs. 10.3 cm/s), the reasons are two fold. (1) Here $v_{geo}$ at point R is essentially an across-shelf velocity component, while that at point Q has both along- and across-shelf components. Note that geostrophic balance is the dominant dynamics in the across-shelf momentum equation ($v = \frac{g}{f}\frac{\partial h}{\partial x}$) on the WFS, but not so in the along-shelf direction where the pressure gradient, surface and bottom stresses are all important (Liu and Weisberg, 2005b). When surface stress (Ekman velocity component) is considered, the standard deviation of ($v_{geo}+v_{Ekm}$) is increased to 9.7 cm/s, closer to that of the observed ($v_{rad}$), 7.9 cm/s (Fig. 9). (2) The quality of coastal altimetry generally decreases towards the coast, and uncertainties in the estimated $v_{geo}$ increase when getting closer to the shore. Adding Ekman component does not reduce the rmsd between the estimated and the observed velocity anomalies or the standard deviation of the velocity difference, because these values are already very low.

## 6. Summary and discussions

Based on multiple-year ocean current observations from a coastal ocean observation system on the West Florida Shelf, the altimetry product provided by the X-TRACK software is assessed for potential usefulness in inferring coastal ocean circulation information over a wide continental shelf. Across-track, surface geostrophic velocity component anomalies, derived from the along-track SLA slope remotely sensed by altimeters onboard the T/P, J1, and T/P interlaced orbit satellite missions are compared with the near surface velocity components from moored ADCP observations at mid shelf (around the 50 m isobath). The altimeter-derived velocity component anomalies are also compared with surface (radial) velocity components by HF radar as sampled perpendicular to the satellite track.

Preliminary results indicate the potential usefulness of the coastal altimetry product, which can be seen from the comparisons of the estimated and observed velocity anomaly time series. The root-mean-square difference (rmsd) between the estimated and the ADCP or HF radar observed near surface velocity component anomalies are 8 – 11 cm/s on subtidal time scales. Given expected velocity errors of 4 – 6 cm/s from the optimal filter (Powell and Leben, 2004), and rmsd of ~ 6 cm/s for deep oceans (e.g., Strub et al., 1997), these 8 – 11 cm/s rmsd values are encouraging and are similar to those found by Le Hénaff et al. (2010) with high resolution coastal altimetry along the continental slope of Spain. This indicates usefulness of the X-TRACK product on the WFS. Note there is an rmsd of 3 – 6 cm/s between the HF radar and ADCP near-surface velocities on subtidal time scales over the WFS (Liu et al., 2010). The smaller rmsd values around the 50 m isobath from HF radar data than the moored ADCP data may be due to better comparability of the HF radar data, i.e., the radar data are measured at surface, and the comparisons are performed at the same points and the same directions (radial current direction is the same as the across-track direction) that velocity projection is not needed. Adding a wind-driven Ekman velocity component generally helps to reduce the rmsd values, as previously found in other regions (e.g., Saraceno et al., 2008).

We should realize that such comparisons are challenging because different instruments measure variables at different scales with different precisions (e.g., Powell et al., 2006). This is particularly vexing because we are trying to estimate and/or observe velocity component that are comparable in magnitude at times to their instrumental errors, given the low energy nature of the WFS (Liu et al., 2010). Also, the satellite tracks are not normal to the direction of the principal axis currents (Liu and Weisberg, 2005b) or the long-term mean currents (Weisberg et al., 2009b) on the shelf, which are mainly in the along-shelf direction. Approaching the coast in the across-shelf direction, the subtidal shelf currents tend to be more polarized in the along-shelf direction (Liu et al., 2007). The large angle between the dominant shelf currents and the altimetry-derived velocity (normal to the satellite track) may affect the velocity comparison. Previous study in the Mediterranean showed that better statistical results of such comparisons (although on 60-day low-pass filtered velocities) were found in the alongshore direction because the main current component and its variability are essentially alongshore (Bouffard et al., 2008). However, our preliminary analysis provides some useful information for the potential applications of altimetry in coastal oceans.

The current version of the X-TRACK product for the WFS area still uses the global models for tides and dynamic atmospheric corrections that have been proven to be efficient over other continental shelves, but may incorrectly represent frictional dissipation due to intrinsic limitations (spatial resolution, non-linear interactions, internal tides parameterization, uncertainties in the atmospheric forcing, etc). Improvements in the velocity comparisons can be expected if more accurate local tidal and atmospheric models are employed in the future.

More interesting work would be an evaluation of the standard along-track altimetry product such as AVISO on the WFS using the same in situ data and procedures. A detailed comparison between the X-TRACK and the AVISO data themselves, and their derived velocity anomalies versus the in situ observations, will be reported in a future paper.

HF radar currents have been used to evaluate the altimeter-derived alongshore currents on the US west coast (Saraceno et al., 2008), but what was used was the "total" velocity that had been combined from two or more radial velocities from different HF radars (e.g., Barrick et al., 1977; Lipa and Barrick, 1983). Note that the accuracy of the "total" velocity is subject to the geometrical dilution of precision (GDOP; e.g., Chapman et al., 1997) owing to the angle at which the radial velocities intersect. Uncertainties are added in combining the radial velocities to form "total" velocities. So, radial velocities themselves may be more useful for the purpose of velocity comparison in this paper. For the same reason, the radial velocities were preferred over the total velocities in data assimilation into a coastal ocean circulation model (Barth et al., 2008). Our long-range HF radar data are gappy due to the low energy nature of the WFS (Liu et al., 2010), and it is difficult to find continuous data that have more overlaps with the satellite altimeter time stamps. As the HF radar systems have been popular around the world's coastal oceans (e.g., Broche et al., 1987; Paduan and Rosenfeld, 1996; Gurgel et al., 1999; Sentchev et al., 2006; Shay et al., 2008; Abascal et al., 2009; Forget et al., 2009; Kaplan et al., 2009; Allou et al., 2010), there will be more opportunities for such velocity comparisons in other regions with better HF radar data returns (e.g., Kosro et al., 1997; Kohut et al., 2006).

In summary, the WFS provides a pilot area for satellite altimetry applications over shelves. The availability of long-term in-situ observations from a number of instruments facilitates intercomnparions. Efforts to improve satellite altimetry in the coastal and shelf seas are being carried out by a lively community at international level (Vignudelli et al., 2011). There is no doubt that new coastal altimetry products will constitute a piece of the coastal observing system puzzle (Cipollini et al., 2010; Vignudelli and Benveniste, 2010). Future altimeter missions (e.g., AltiKa, Sentinel-3, SWOT) carrying a new generation of altimeters also promise to enhance resolution and extend capabilities as close as possible to the coast. In any scenario it is evident that the time-space density of altimeter-derived measurements would be more complete. The use of these observations in synergy with other data sources and modeling tools will enable a better understanding and prediction of the structure and evolution of the currents in WFS.

## 7. Acknowledgements

Support was by the Office of Naval Research, Grants N00014-05-1-0483 and N00014-10-0785 and by NOAA Grants NA06NOS4780246 and NA08NOS4730409. The first two are for observing and modeling thee West Florida Shelf circulation. The third is for applications to harmful algae under the ECOHAB program, and the fourth, via SC SeaGrant, is for maintaining high-frequency radars for SECOORA as part of the NOAA-IOOS Program. The HF radar-processing toolbox HFR_Progs was provided by D. Kaplan, M. Cook, D. Atwater, and J. F. González. The Venice site was jointly maintained by Mote Marine Laboratory (B. Pederson and G. Kirkpatrick) and USF (C. Merz), with Rutgers University (H. Roarty and S. Glenn) providing the site equipment. The altimeter products were produced and distributed by Aviso (http://www.aviso.oceanobs.com/), as part of the Ssalto ground processing segment. Altimetry data used in this study were developed, validated, and distributed by the CTOH/LEGOS, France. Financial support by CNES is acknowledged (MARINA OST-ST project). This is CPR contribution 18.

Table 1. Statistics of the comparisons between altimetry-derived and observed velocities. The values in the parentheses are calculated with 18 hour time lag between the altimetry-derived and the observed velocities.

| Satellite & track # | Velocity obs. | Velocity component | Time period | Data points | Standard deviation | | | rmsd | |
|---|---|---|---|---|---|---|---|---|---|
| | | | | | $V_{obs}$ | $V_{geo}$ | $V_{geo} + V_{ekm}$ | $V_{geo}$ vs. $V_{obs}$ | $V_{geo}+V_{ekm}$ vs. $V_{obs}$ |
| T/P #091 | mooring PM1 | Normal | 30-Sep-1993 ~31-Jan-1995 | 41 | 7.0 | 8.1 | 8.8 | 8.6 | 8.4 |
| T/P #091 | mooring PM1 | along-shelf | 30-Sep-1993 ~31-Jan-1995 | 40 | 9.0 | 9.9 | 10.8 | 9.9 | 9.8 |
| J1 #091 | mooring C12 | along-shelf | 01-Jan-2002 ~01-Feb-2008 | 75 | 8.4 | 9.9 | 11.6 | 9.9 (9.0) | 9.7 (9.3) |
| J1 #091 | mooring C13 | along-shelf | 01-Jan-2002 ~01-Feb-2008 | 169 | 7.5 | 9.1 | 11.5 | 9.9 (10.1) | 10.5 (10.5) |
| T/P interleaved #167 | HF radar | radial Q | 22-Apr-2004 ~06-Oct-2005 | 23 | 4.9 | 10.3 | 9.3 | 11.1 | 10.3 |
| T/P interleaved #102 | HF radar | radial R | 22-Apr-2004 ~06-Oct-2005 | 23 | 7.9 | 3.8 | 9.7 | 8.2 | 9.3 |

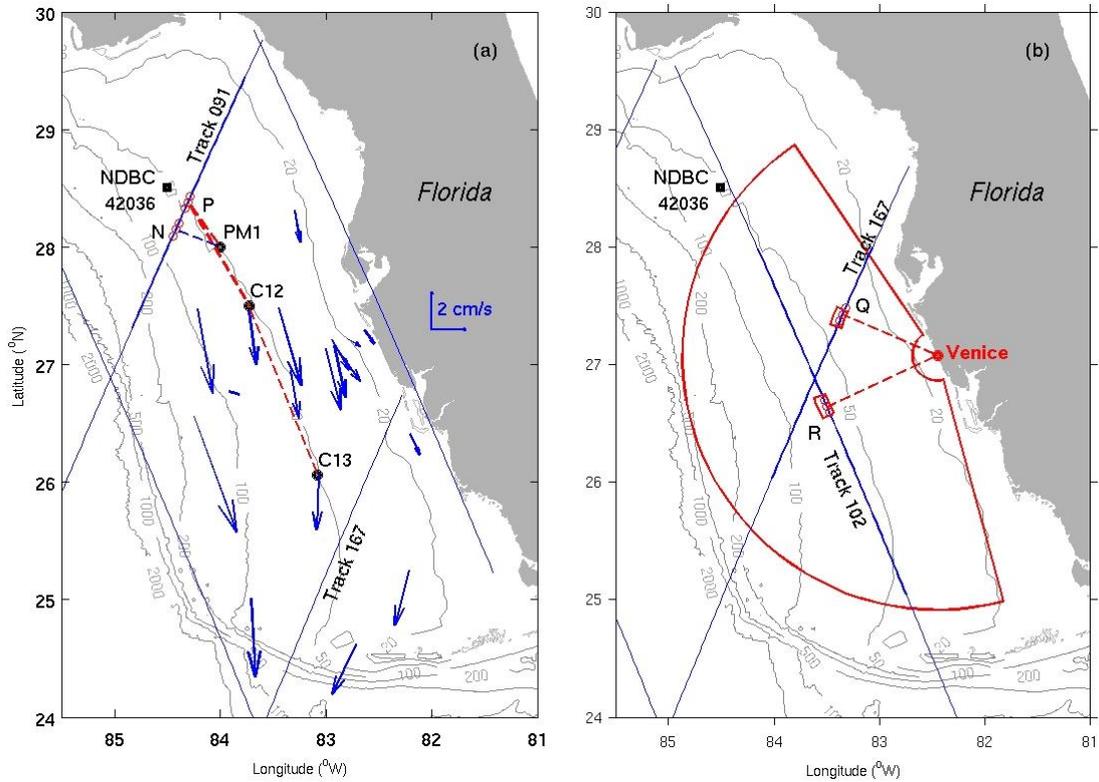

Fig. 1. Altimetry satellite tracks, mooring locations, and HF radar spatial coverage on the West Florida Shelf. (a) J1 and T/P satellite ground tracks (blue lines) and ADCP mooring locations (PM1, C12 & C13). The long-term mean depth-averaged currents (Weisberg et al., 2009b) are shown as vectors. (b) T/P Interleaved satellite ground tracks and HF radar coverage map (red arcs). NOAA/NDBC Buoy 42036 is shown on both maps. Also shown are bathymetry contours (units in m).

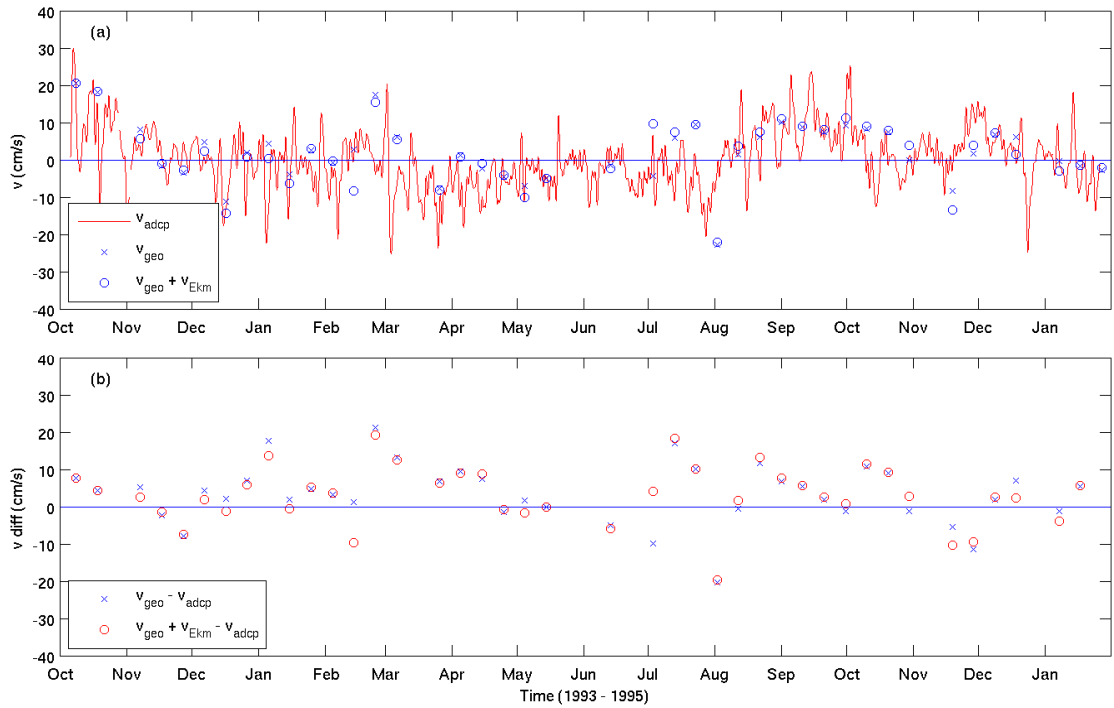

Fig. 2. Comparison of velocity components perpendicular to the T/P satellite track 091 (Points N and PM1 in Fig. 1a): surface geostrophic velocity calculated from along-track altimetry (+ surface Ekman velocity) versus ADCP near-surface (4 m) velocity anomalies (a) and their differences (b). Both the ADCP and wind time series are 36 hour lowpass filtered.

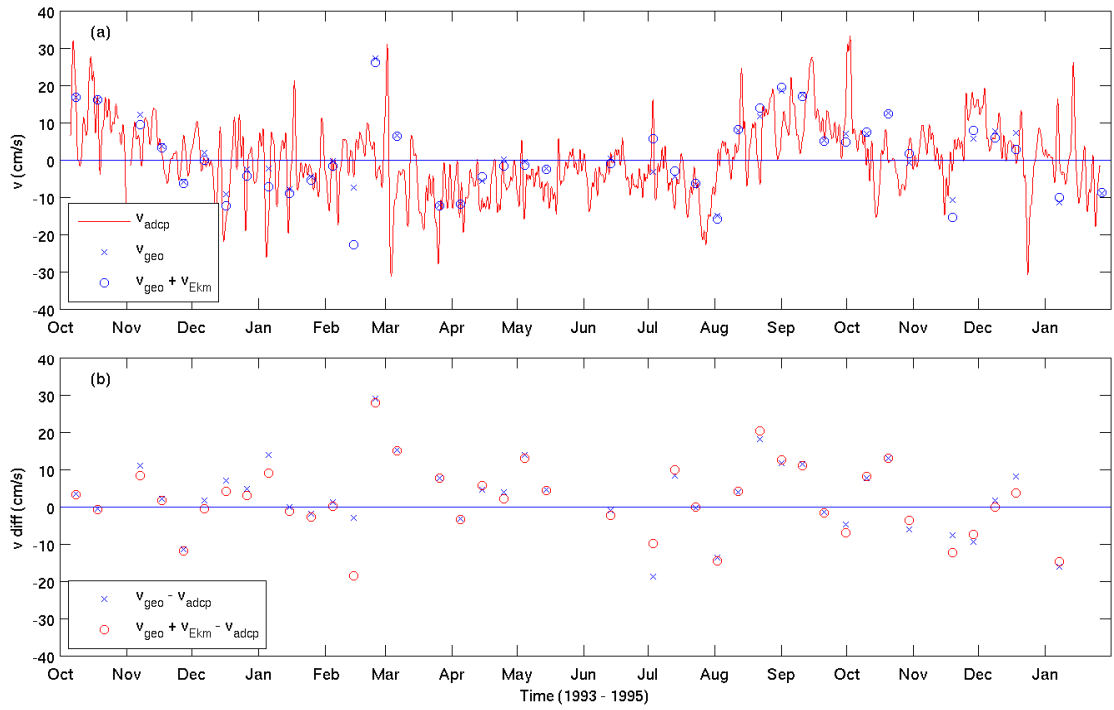

Fig. 3. Same as Fig. 2 but for along-shelf component (line P – PM1 in Fig. 1a).

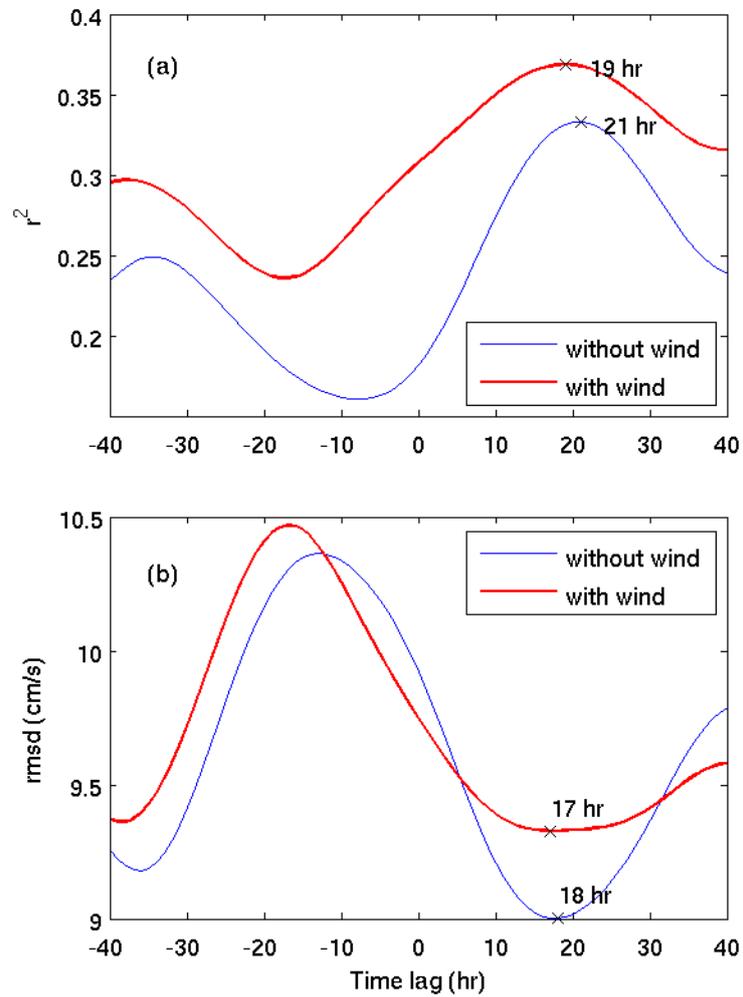

Fig. 4. Time-lagged (a) correlation coefficient ($\gamma^2$) and (b) root-mean-squared difference (rmsd) between the altimetry-derived and observed along-shelf velocity at mooring C12. Positive time lag means moored velocity lags the altimetry-derived velocity. The time lags corresponding to the maximum $\gamma^2$ and the minimum rmsd values are shown.

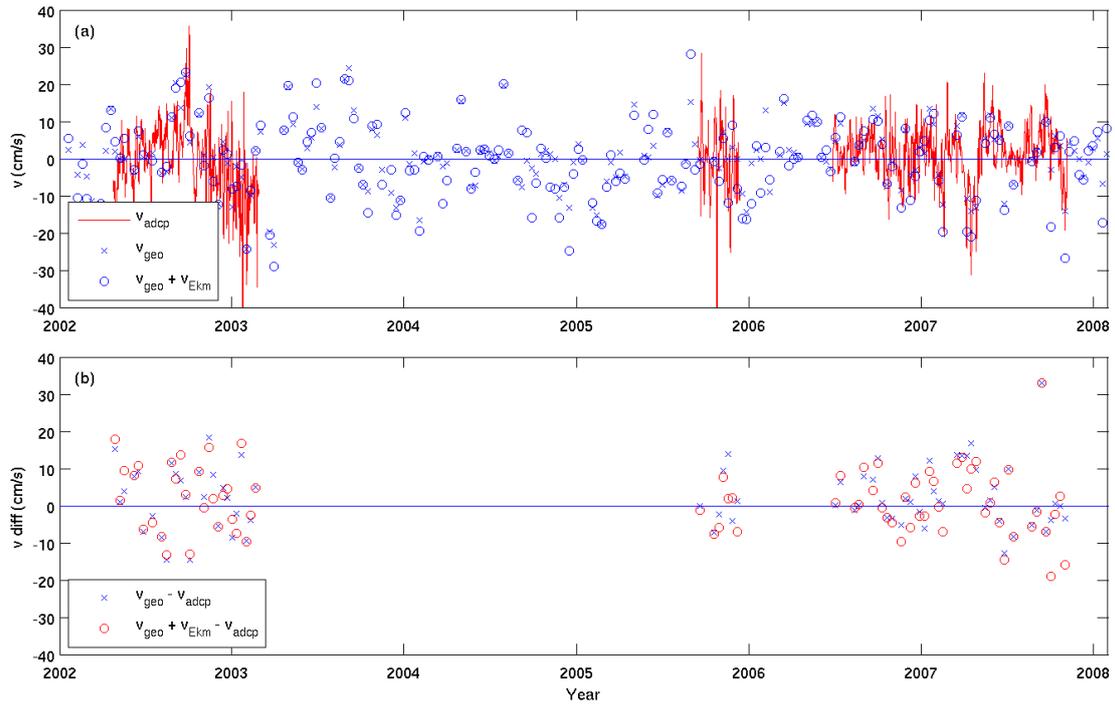

Fig. 5. Comparison of along-shelf velocity components at the 50 m isobath: surface geostrophic velocity calculated from J1 altimetry along track 091 (+ surface Ekman velocity) versus ADCP near-surface (4 m) velocity anomalies at mooring C12 (a) and their differences (b). Both the ADCP and wind time series are 36 hour lowpass filtered.

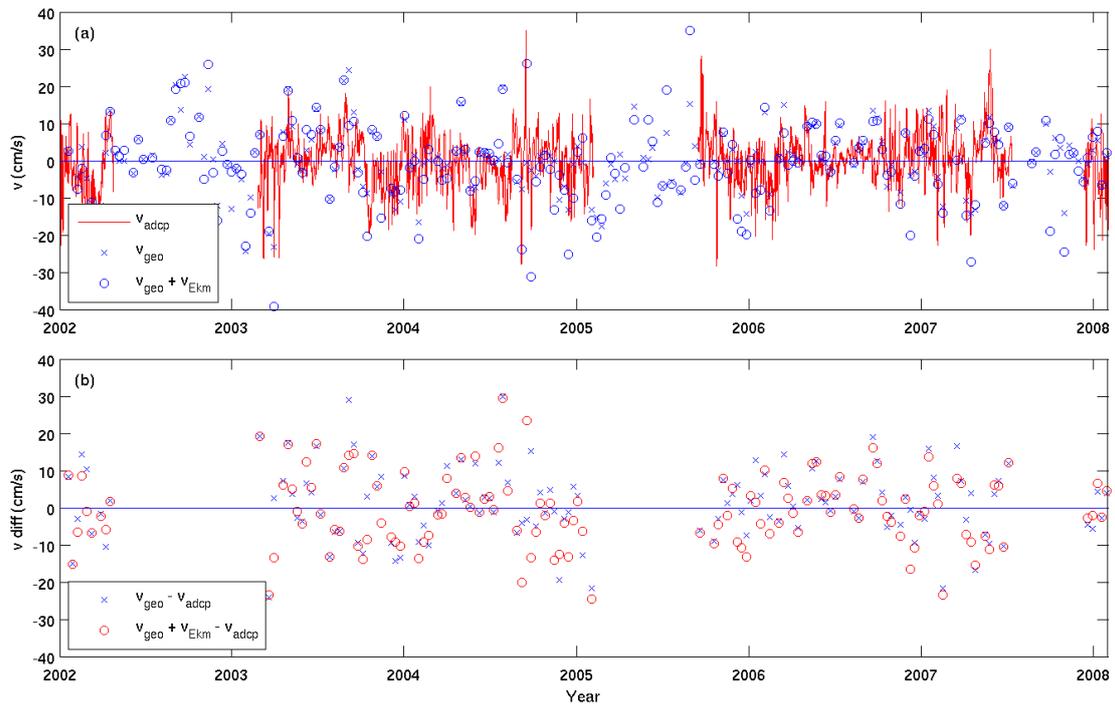

Fig. 6.  Same as Fig. 5 but for mooring C13.

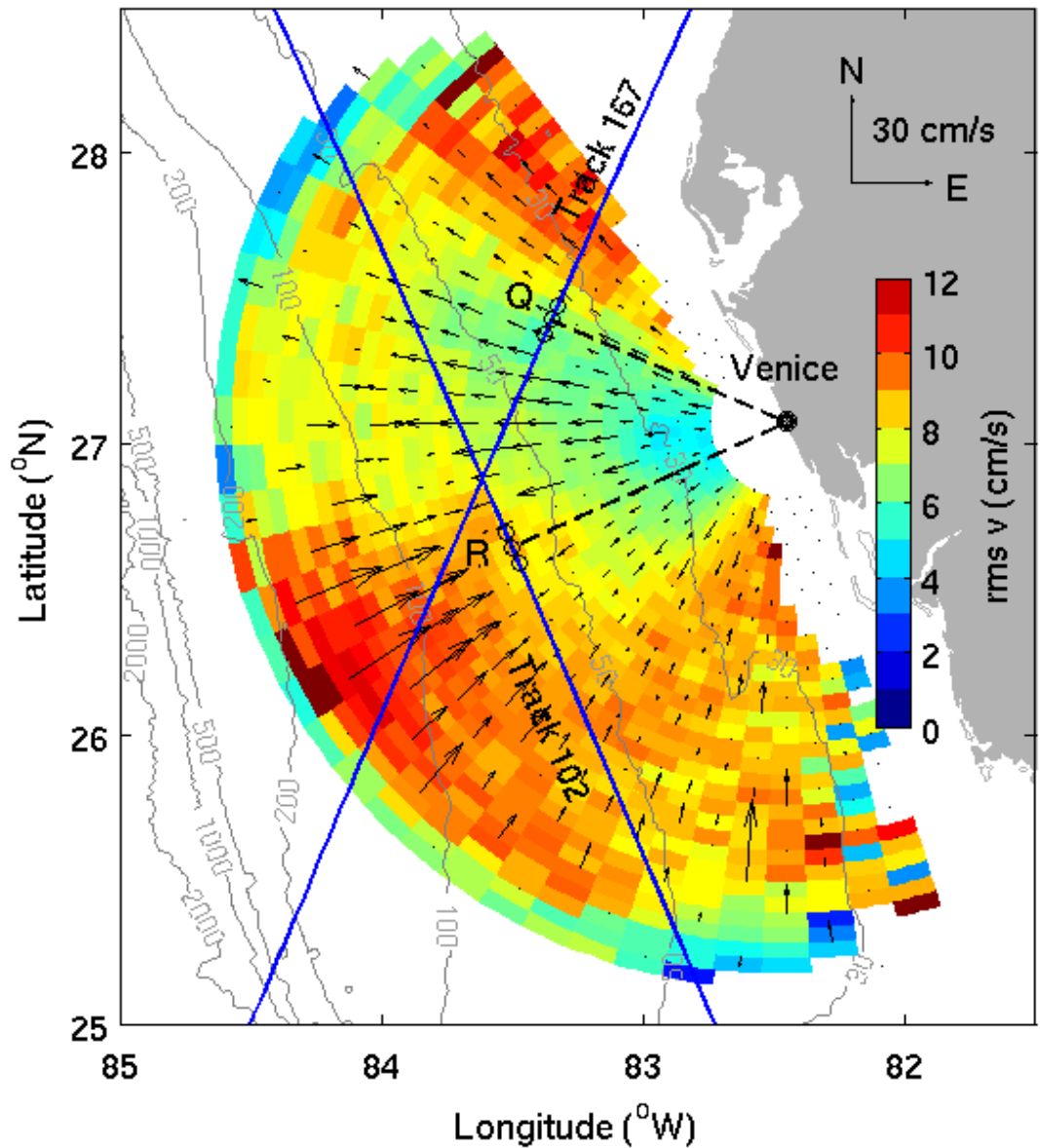

Fig. 7. A snapshot of HF radar radial velocity field superimposed on the root-mean-squared (rms) values of the 30 hour lowpass filtered radial currents (color) as measured from the Venice site. The radial velocities are subsampled for clarity. Bathymetry units in meters. Also shown are the satellite ground tracks from the T/P interleaved mission, and the two points (Q and R) where the across-track directions are the same as the radial current directions.

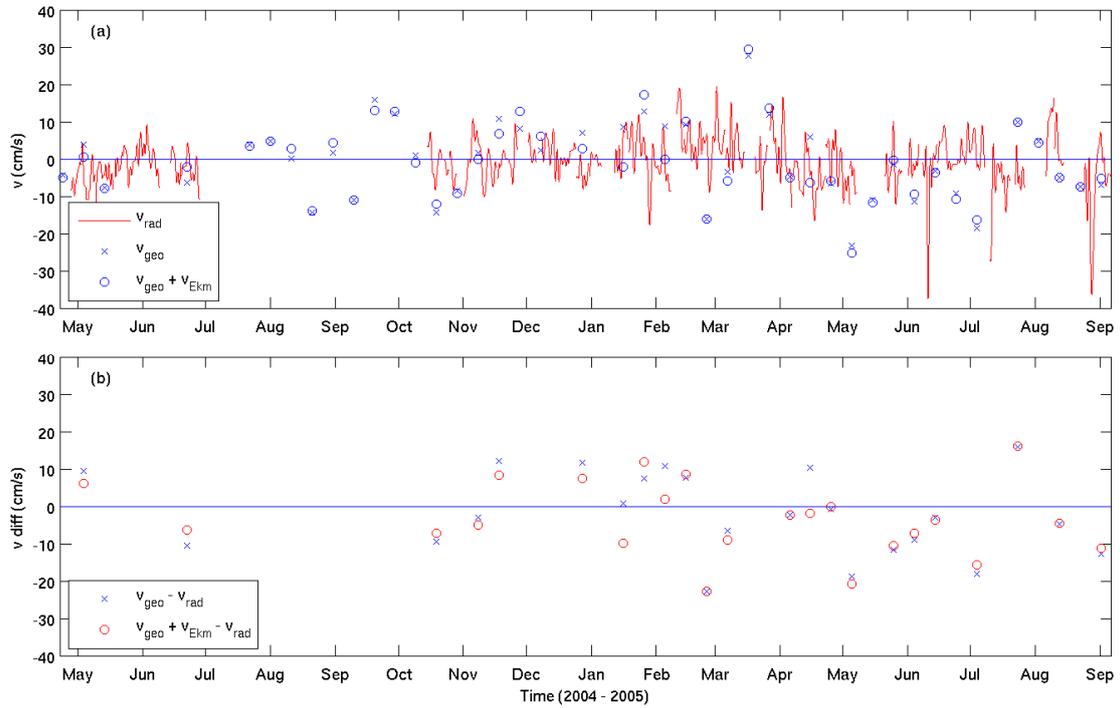

Fig. 8. Comparison of velocity components perpendicular to T/P Interleaved ground track 167 at Point Q in Figure 7: surface geostrophic velocity calculated from T/P Interleaved altimetry along track 167 (+ surface Ekman velocity) versus HF radar radial velocity anomalies (a) and their differences (b). Both the current and wind time series are 30 hour lowpass filtered.

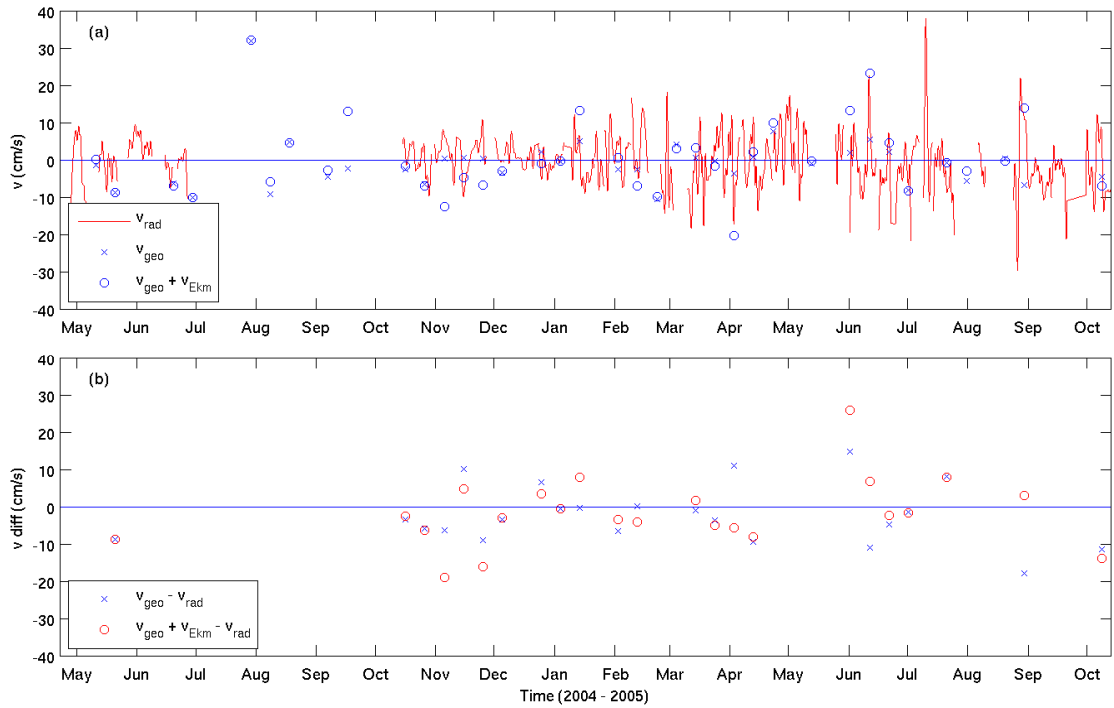

Fig. 9. Comparison of velocity components perpendicular to T/P Interleaved ground track 102 at Point R in Figure 7: surface geostrophic velocity calculated from T/P Interleaved altimetry along track 102 (+ surface Ekman velocity) versus HF radar radial velocity anomalies (a) and their differences (b). Both the current and wind time series are 30 hour lowpass filtered.